# PM$_{2.5}$ as a major predictor of COVID-19 basic reproduction number in the USA

**Ognjen Milicevic[1], Igor Salom[2], Andjela Rodic[3], Sofija Markovic[3], Marko Tumbas[3], Dusan Zigic[2], Magdalena Djordjevic[2], Marko Djordjevic[3,*]**

[1]Department for Medical Statistics and Informatics, School of Medicine, University of Belgrade, Serbia

[2]Institute of Physics Belgrade, National Institute of the Republic of Serbia, University of Belgrade, Serbia

[3]Quantitative Biology Group, Institute of Physiology and Biochemistry, Faculty of Biology, University of Belgrade, Serbia

**\* Correspondence:**

Marko Djordjevic, e-mail: dmarko@bio.bg.ac.rs

**Keywords:** COVID-19 pollution dependence, outdoor air pollutants, basic reproduction number, principal component analysis, machine learning

## Abstract

Many studies have proposed a relationship between COVID-19 transmissibility and ambient pollution levels. However, a major limitation in establishing such associations is to adequately account for complex disease dynamics, influenced by e.g. significant differences in control measures and testing policies. Another difficulty is appropriately controlling the effects of other potentially important factors, due to both their mutual correlations and a limited dataset. To overcome these difficulties, we will here use the basic reproduction number ($R_0$) that we estimate for USA states using non-linear dynamics methods. To account for a large number of predictors (many of which are mutually strongly correlated), combined with a limited dataset, we employ machine-learning methods. Specifically, to reduce dimensionality without complicating the variable interpretation, we employ Principal Component Analysis on subsets of mutually related (and correlated) predictors. Methods that allow feature (predictor) selection, and ranking their importance, are then used, including both linear regressions with regularization and feature selection (Lasso and Elastic Net) and non-parametric methods based on ensembles of weak-learners (Random Forest and Gradient Boost). Through these substantially different approaches, we robustly obtain that PM$_{2.5}$ is a major predictor of $R_0$ in USA states, with corrections from factors such as other pollutants, prosperity measures, population density, chronic disease levels, and possibly racial composition. As a rough magnitude estimate, we obtain that a relative change in $R_0$, with variations in pollution levels observed in the USA, is typically ~30%, which further underscores the importance of pollution in COVID-19 transmissibility.

## 1    Introduction

In the current era of globalization, the appearance of the new SARS-CoV-2 virus in 2019 has harshly reminded humanity of how easily an epidemic can also become a global issue. While essentially the entire world, already for more than a year, suffers from the COVID-19 disease, not all areas have been hit equally. Hence, scientists worldwide are struggling to find patterns in observable variations in the epidemic progression speed and/or its severity, and the present paper is a part of this international and interdisciplinary effort (Bontempi et al., 2020). More specifically, we aim to understand the possible effects of air pollution on the transmission of COVID-19.

Many previous studies have already provided arguments for the importance of pollution (primarily $PM_{2.5}$ and, to a lesser degree, $PM_{10}$ and $NO_2$) in COVID-19 transmissibility and suggested mechanisms that might explain this connection. It was argued that droplets with virus particles may bind to Particulate Matter (PM), which may promote the diffusion of virus droplets in the air (Chen et al., 2010; Comunian et al., 2020; Contini and Costabile, 2020). Furthermore, once the virus droplet bound to PM reaches a susceptible individual, it can penetrate deeper in alveolar and tracheobronchial regions – especially in the case of small ($PM_{2.5}$) pollution particles (Copat et al., 2020; Qu et al., 2020). Besides these direct mechanical effects on transmission, pollution has a general effect on weakening the immune system making the organism more susceptible to infection (Domingo and Rovira, 2020; Paital and Agrawal, 2020; Qu et al., 2020). In addition, it promotes overexpression of ACE-2 receptors, which allows SARS-CoV-2 binding and entry into cells (Comunian et al., 2020; Paital and Agrawal, 2020; Sagawa et al., 2021).

While these arguments are compelling, and several studies pointed to correlations between pollutant levels and increased severity of COVID-19 progression (De Angelis et al., 2021; Kolluru et al., 2021; Lorenzo et al., 2021; Tello-Leal and Macías-Hernández, 2020; Yao et al., 2021; Zhu et al., 2020), there are also prominent methodological difficulties in establishing this link, as discussed in (Anand et al., 2021; Bontempi, 2021; Bontempi et al., 2020; Villeneuve Paul J. and Goldberg Mark S., 2020). Specifically, comparing case counts (Adhikari and Yin, 2020; Suhaimi et al., 2020) in different geographical regions may be influenced by significant differences in the epidemic onsets (Villeneuve Paul J. and Goldberg Mark S., 2020), applied control (e.g., social distancing) measures (Bontempi, 2021)), and testing methodologies (most significantly the number of performed tests). Consequently, adequately controlling for the infection dynamics, rather than relying on absolute case counts, is crucial. Secondly, due to a multitude of potential confounding factors, it is crucial to, jointly with pollution, consider possible influences of diverse sociodemographic, economic, medical, and meteorological factors on transmission (Bontempi, 2020b; Bontempi et al., 2020). Ideally, the scope of the study should be conceived to emphasize variability in pollution, while being relatively homogenous in these other factors. As another obstacle, the considered variables can be mutually highly correlated (Notari and Torrieri, 2021; Salom et al., 2021). Such high correlations realistically present a problem for any statistical inference method, though modern machine-learning approaches can partially account for this difficulty (Gupta and Gharehgozli, 2020). Additionally, the relationship of input variables to $R_0$ might be (highly) non-linear, which can hardly be accounted for by linear regressions, but may be successfully addressed by e.g. ensembles of decision trees (Hastie et al., 2009). Finally, to obtain robust predictions that are not an artifact of the applied methodology and the underlying assumptions, it is crucial to perform analysis by several independent methods.

In our approach, we aim to address these general limitations. First of all, the USA dataset seems to be optimal for this analysis: while, in absolute figures, the pollution in the United States is not high, there is still sufficient variability in the pollution variables to extract reasonable conclusions, whereas heterogeneities in sociodemographic and weather parameters are not too large to overshadow the dependence on pollution. Next, as a measure of transmissibility, we use the basic reproduction number ($R_0$). $R_0$ is a measure of SARS-CoV-2 transmissibility in a completely susceptible (non-resistant) population and in the absence of social distancing (sometimes also referred to as $R_{0,\text{free}}$ (Magdalena Djordjevic et al., 2021b; Maier and Brockmann, 2020)), which is insensitive to differences in specific testing policies and control measures. We here apply our previously developed methodology (Salom et al., 2021), which is based on observation of different dynamical regimes in COVID-19 infection counts during the disease outburst (Magdalena Djordjevic et al., 2021a). Our model is then applied to one of these growth regimes (the exponential one), to estimate $R_0$ for individual USA states. These $R_0$ estimates, instead of the disease counts (or other similar measures), are then used as the dependent (response) variable in further analysis. As independent (input) variables, we assemble a large set of available sociodemographic, medical, and weather variables. Importantly, to assess the pollution levels in detail, we assemble the data for ten different pollutants, with the levels determined in the time windows relevant for the analyzed exponential growth regimes. We gather the weather parameters in the same dynamically relevant manner. This results in a large number of predictors, many of which we group in sets of similar and mutually tight highly correlated variables. Additionally, the number of assembled variables exceeds the total sample size, so it is necessary to reduce the number of predictors to a smaller and less correlated set. We achieve this through data preprocessing (feature engineering), which includes variable transformations, removing all outliers, and grouping mutually related and highly correlated variables into subsets (e.g., age-related, population prosperity measures, chronic diseases). Principal Component Analysis (PCA) is then applied within these subsets, resulting in dimensionality reduction (reducing the number of predictors) and smaller overall correlations within this reduced predictor set. Finally, to go beyond establishing mere correlations between different variables/components with R0, we use four established machine learning approaches: Lasso, Elastic net, Random Forest, and Gradient Boost. Our goal is to: *i)* select important variables and rank their relative importance in explaining $R_0$, *ii)* obtain an estimate of expected changes in $R_0$ based on observed variability in pollution levels. While the estimates we get in this way are only rough (due to the inability to assemble all relevant factors in determining $R_0$), the obtained results nevertheless provide a quantitative assessment of the importance of pollution in SARS-CoV-2 transmissibility.

## 2    Methods

### 2.1 $R_0$ extraction

As the proxy for the COVID-19 transmissibility, we used the basic reproduction number ($R_0$). Basic reproduction number is a measure of SARS-CoV-2 transmissibility in a fully susceptible population and in the absence of intervention measures (social distancing, quarantine). For extraction of $R_0$, we used our previously published methodology, in particular analysis of widespread infection growth regimes (Magdalena Djordjevic et al., 2021a) and extraction of $R_0$ from the exponential growth phase that we previously applied on a worldwide level (Salom et al., 2021). For the sake of completeness, we summarize this methodology below.

To describe the SARS-CoV-2 transmission in a population, we constructed an adapted version of an SEIR compartmental model (Maier and Brockmann, 2020; Maslov and Goldenfeld, 2020; Perkins and España, 2020; Tian et al., 2020; Weitz et al., 2020), which takes into account all the relevant features of this process, while being simple enough to be used for $R_0$ estimation in a wide range of populations (Magdalena Djordjevic et al., 2021a; Salom et al., 2021). In the early stages of epidemics and before social distancing measures are introduced, the flow between the model compartments leads to the changes of the compartment member abundances $S$ (susceptible), $E$ (exposed), $I$ (infected), $R$ (recovered), and $D$ (cumulative detected cases) which are described by the following system of ordinary differential equations:

$$\frac{dS}{dt} = -\frac{\beta SI}{N} \tag{1.1.}$$

$$\frac{dE}{dt} = \frac{\beta SI}{N} - \sigma E \tag{1.2.}$$

$$\frac{dI}{dt} = \sigma E - \gamma I \tag{1.3.}$$

$$\frac{dR}{dt} = \gamma I \tag{1.4.}$$

$$\frac{dD}{dt} = \varepsilon \delta I \tag{1.5.}$$

where $N$ is the population size. Parameters represent: $\beta$ - the rate of virus transmission from an infected to the encountered susceptible individual, $\sigma$ - the inverse of the average incubation period ($\sim$3 days), $\gamma$ - the inverse of the average period of infectiousness, $\varepsilon$ – the detection efficiency (as not every infected individual becomes detected), and $\delta$ - the detection rate.

We here applied the model to the relatively brief, initial epidemics period when only a small fraction of the population is resistant, and before social distancing interventions take effect. Note that, even after introducing the measures, there is $\sim$10 days delay in observing their effect in the confirmed case-counts curve, due to the incubation period and the time needed between the symptom onset and the infection detection/confirmation. During this period, the virus is spreading at a rate determined by its natural biological potential, modulated by the characteristics of the given population and the environment. Therefore, the above parameter values of infection progression are considered constant in this period. The standard measure of the virus transmissibility in these conditions (not influenced by interventions or immunity) is the basic reproduction number, $R_0$, defined as the average number of secondary infections caused by a primary infected individual in a fully susceptible population ($S/N \approx$ 1), and in the absence of social distancing measures (also sometimes denoted as $R_{0,free}$) (Maier and Brockmann, 2020). At the start of an epidemic, $R_0 > 1$ and the number of infected individuals grows exponentially. The model can then be linearized by invoking $S/N \approx 1$, reducing the model to two linear differential equations (1.2) and (1.3). Solving for the eigenvalues of this system,

$$\lambda_{\pm} = \frac{-(\gamma + \sigma) \pm \sqrt{(\gamma - \sigma)^2 + 4\beta\sigma}}{2}, \tag{1.6.}$$

provides the solution of the form $I(t) = C_1 \cdot e^{\lambda_+ t} + C_2 \cdot e^{\lambda_- t}$, which can be approximated by

$$I(\text{t}) = I(0) \cdot e^{\lambda_+ t} \tag{1.7.}$$

where the term containing the negative eigenvalue, $\lambda_-$ can be neglected (see (Salom et al., 2021)). With $R_0 = \beta/\gamma$ (Keeling and Rohani, 2011; Martcheva, 2015), the equation for the basic reproduction number,

$$R_0 = 1 + \frac{\lambda_+ \cdot (\gamma + \sigma) + \lambda_+^2}{\gamma * \sigma}. \tag{1.8.}$$

can be obtained by expressing $\beta$ from Eq. (1.6).

To estimate the $R_0$ values for 46 US states, we collect the detected case counts for each state from online resources (Worldometer, 2020). The solution $D(\text{t}) = \varepsilon \cdot \delta \cdot I(0) \cdot (e^{\lambda_+ t} - 1)/\lambda_+$ of the Eq. (1.5) using the Eq. (1.7) models the dependence of the cumulative number of detected with time. Taking its logarithm

$$\log(D(\text{t})) = \log\left(\varepsilon \delta I(0)/\lambda_+\right) + \lambda_+ \cdot \text{t}, \tag{1.9.}$$

results in the equation of the straight line that can be fitted to the data on the semilogarithmic scale. Notably, the slope of that line is given by the positive eigenvalue of the system, $\lambda_+$. Once that $\lambda_+$ is determined by fitting, the value of $R_0$ for a particular state can be calculated from Eq. (1.8).

## 2.2 Pollution data collection

Air quality information was obtained from the US environmental protection agency (EPA) Air Data service (US Environmental Protection Agency, 2020). We used aggregated daily data for pollutant gases ($O_3$, $NO_2$, $SO_2$, CO), particulates ($PM_{2.5}$ and $PM_{10}$) and other available species, such as VOCs (Volatile Organic Compounds), NOx, and HAPs (Hazardous Air Pollutants). For a given state, aggregation was done over all cities with available information. The populations of cities were obtained from the US Census Bureau (U.S. Census Bureau, 2020). All the variable values are averaged for each city over the identified time period, and the state average is calculated as the average of all included state cities weighted by the population.

## 2.3     Weather data collection

Weather parameters were downloaded in bulk using a custom Python script from the NASA POWER project service (NASA Langley Research Center, 2020). All the parameters were downloaded via the POWER API at the longitude and latitude coordinates matching the largest cities in each state that comprise above 10% of the state population. Variables include temperature at 2m and 10m, measures of humidity and precipitation (wet bulb temperature, relative humidity, total precipitation), and insolation indices. The maximum predicted UV index was downloaded from OpenUV (OpenUV, 2020). Geographical coordinates of the cities and populations of cities and states were adapted from Wikidata (Wikipedia, 2021a, 2021b).

## 2.4 Socio-demographic data collection

Demographic data were collected from several sources. The demographic composition of the US population by gender, race, and percentage of the population under 18 and over 65 was taken from the Measure of America, a project of The Social Science Research Council website (Measure of America, 2018). Information about health insurance, GDP, life expectancy at birth, infant and child mortality was also taken from the Measure of America website. Medical parameters such as hypertension, cholesterol, cardiovascular disease, diabetes, cancer, obesity, inactivity, and chronic kidney and obstructive pulmonary disease were taken from America's Health Rankings website (America's Health Ranking, 2021) hosting Centers for Disease Control and Prevention (CDC) data (CDC, 2019). Percentages of the population that are actively smoking and consuming alcohol are taken from the same source. The percentage of the foreign population was taken from the Census Reporter website (U.S. Census Bureau, 2019). The subnational HDI was taken from the Global Data Lab website (2020) (Smits and Permanyer, 2019). Population density, urban population percentage, and median age were taken from the U.S. Census Bureau website (U.S. Census Bureau, Population Division, 2019).

## 2.5     Data processing

The initial analysis of the assembled data distributions and QQ plots revealed non-normal distributions in a majority of variables. To reduce the skewness of the data we applied a number of transforms with different strengths (square root, cubic root, or log), adjusted in sign to maintain the data ranking (Spearman correlation). Individual data values that remained more than three median absolute deviations from the new median were substituted by the said median value.

The main purpose of these transformations, and outliers' removal, was to account for more extreme variable values (such as heavy distribution tails), which may significantly affect some of the analysis methods that we further use (in particular, correlation analysis, Lasso and Elastic net regressions). On the other hand, methods based on the ensembles of decision trees (e.g., Random Forest and Gradient Boost) are fairly robust to outliers and non-normal variable distributions and provide a consistency check of the obtained conclusions.

The table with all applied transformations is provided below. Also, note that the entire dataset used in this analysis (variable values for all 46 states) is provided in Supplement Table 1. In addition to the transformations applied, the table below also links the variables to the dataset, by relating a variable shortcut (used in Supplemental files) with its full name and units.

| Data | Name (units) | Transformation f(x) |
|---|---|---|
| T2M, T2M$_{MAX}$, T2M$_{MIN}$, T10M, T10M$_{MAX}$, T10M$_{MIN}$, TS, T2MWET | Temperatures (°C) | None |
| RH2M | Relative humidity at 2 meters (%) | $-\log(\max(x) - x)$ |
| QV2M | Specific humidity at 2 meters (g/kg) | $\log(x)$ |
| T2MDEW | Dew Point (°C) | None |
| PRECTOT | Precipitation (mm/day) | $x^{1/3}$ |
| TQV | Total Column Precipitable Water (cm) | $\log(x)$ |
| CLRSKY_SFC_SW_DWN | Clear Sky Insolation Incident on a Horizontal Surface (MJ/m$^2$/day) | $-(\max(x) - x)^{1/3}$ |
| ALLSKY_SFC_LW_DWN | Downward Thermal Infrared (Longwave) Radiative Flux (MJ/m$^2$/day) | $\log(x)$ |
| ALLSKY_SFC_SW_DWN | All Sky Insolation Incident on a Horizontal Surface (MJ/m$^2$/day) | $\log(x)$ |
| OpenUV$_{max}$ | UV radiation index | $x^{1/3}$ |
| WS2M | Wind speed at 2 meters | None |
| WS10M | Wind speed at 10 meters | None |
| P | Pressure | $x^{1/2}$ |
| Population over 65 (%) | Population over 65 (%) | None |
| Life Expectancy | Life Expectancy at Birth (years) | $-(\max(x) - x)^{1/2}$ |
| Median age | Median age (years) | $-(\max(x) - x)^{1/2}$ |
| Youth population | Population under 18 (%) | $\log(x)$ |
| Population density | Population density (people/km$^2$) | $\log(x)$ |
| BUAPC | Built Up Area Per Capita (km$^2$/people) | $\log(x)$ |
| Urban Population | Urban Population (%) | $-(\max(x) - x)^{1/2}$ |
| HDI | Human development index (0-1) Average of education, health and standard of living. (Mean years of schooling of adults aged 25+, Expected years of schooling | $-(\max(x) - x)^{1/2}$ |
| | of children aged 6 + Life expectancy at birth + GNIpc ) /3 | |
| GDPpc | Gross domestic product per capita | log(x) |
| Infant mortality rate | Infant Mortality Rate (per 1,000 live births) | -log(x) |
| Child mortality | Child Mortality (age 1-4, per 1000 population) | -log(x) |
| Alcohol consumption | Adults alcohol consumption binge drinking (%) | log(x) |
| Foreign-born population | Foreign-born population (%) | log(x) |
| Obesity | Obesity age 20 and older (%) | None |
| CVD deaths | Age 65+ Cardiovascular disease deaths per 100000 people | log(x) |
| Hypertension | Adults with Hypertension (%) | log(x) |
| High cholesterol | Population with high cholesterol (%) | None |
| Smoking | Population smoking (%) | None |
| Cardiovascular disease | Population with cardiovascular disease (%) | None |
| Diabetes | Population with diabetes (%) | $x^{1/3}$ |
| Cancer | Population with cancer (%) | None |
| Chronic kidney disease | Population with chronic kidney disease (%) | $x^{1/2}$ |
| Chronic obstructive pulmonary disease | Population with chronic obstructive pulmonary disease (%) | log(x) |
| Multiple chronic conditions | Population with multiple chronic conditions (%) | None |
| Physical inactivity | Population physically inactive (%) | $x^{1/3}$ |
| Male percent | Fraction of male in the population (%) | log(x) |
| White percent | Fraction of white in the population (%) | -log(max(x) - x) |
| Black percent | Fraction of black in the population (%) | $x^{1/3}$ |
| Native percent | Fraction of native in the population (%) | log(x) |
| Asian percent | Fraction of Asian in the population (%) | log(x) |
| Latino percent | Fraction of Latino in the population (%) | log(x) |
| No health insurance children | No health insurance under 18 (%) | $x^{1/2}$ |
| No health insurance adults | No health insurance 18-64 (%) | None |
| No health insurance all | No health insurance all population (%) | None |
| No insurance black | No health insurance black (%) | None |
| No insurance native | No health insurance native (%) | $x^{1/3}$ |
| No insurance Asian | No health insurance Asian (%) | $x^{1/2}$ |
| No insurance Latino | No health insurance Latino (%) | None |

| No insurance white | No health insurance white (%) | None |
|---|---|---|
| PM$_{2.5}$ | PM$_{2.5}$ concentration (µg/m$^3$) | None |
| PM$_{10}$ | PM$_{10}$ concentration (µg/m$^3$) | $x^{1/2}$ |
| CO | CO concentration (ppm, 10$^{-6}$) | $x^{1/2}$ |
| NO$_2$ | NO$_2$ concentration (ppb, 10$^{-9}$) | None |
| SO$_2$ | SO$_2$ concentration (ppb) | $\log(x - \min(x))$ |
| O$_3$ | O$_3$ concentration (ppm) | None |
| VOC | Volatile organic compounds concentration (ppb Carbon) | $\log(x)$ |
| Lead | Lead concentration (µg/m$^3$) | $\log(x)$ |
| HAPs | Hazardous air pollutants concentration (µg/m$^3$) | $(x-\min(x))^{1/2}$ |
| NONOxNOy | Nitrous oxides concentration (ppb) | $x^{1/3}$ |
| $R_0$ | Estimated basic reproduction number | $\log(x)$ |

**Table 1**. List of variables (with units) and the applied transformations. Variable shortcuts (first column) correspond to Supplement Table 1.

### 2.6 Feature engineering and Principal Components Analysis

The total number of variables (74) is larger than the sample size (46 states). While the regressions with feature selection (Lasso and Elastic net) can handle the number of variables that is significantly larger than the sample size (as long as the number of selected features is smaller than the sample size), this large number of variables (some highly correlated) is a major risk for overfitting, particularly for Random Forest and Gradient Boost methods. To reduce the number of variables, we first divided them into groups by conceptual similarity and expected correlation, after which we performed Principal Component Analysis (PCA) on each group. This also partially reduced data correlation (Joliffe, 2002). Variables were grouped according to two criteria: *i)* those that represent similar quantities so that, after PCA, the interpretation of the obtained PC remains unambiguous; *ii)* the correlations between the variables in the same group are high, so that in this way, after PCA, the overall correlations in the new predictor set are substantially reduced. Grouping of variables and their relation to PCA is provided in Table 2.

| PC components | Variables |
|---|---|
| PC1 temperature | T2M, T2M$_{MAX}$, T2M$_{MIN}$, T10M, T10M$_{MAX}$, T10M$_{MIN}$, TS |
| PC1 humidity | QV2M, T2MDEW |
| PC1 percipitation | PRECTOT, TQV |
| PC1 radiation<br>PC2 radiation | CLRSKY_SFC_SW_DWN,<br>ALLSKY_SFC_SW_DWN, ALLSKY_SFC_LW_DWN |
| PC1 seasonality<br>PC2 seasonality | PC1 T, PC1 humidity, PC1 precipitation, PC1 radiation, PC2 radiation, RH2M, UV |
| PC1 age<br>PC2 age | Population over 65, Youth population, Median age |
| PC1 density<br>PC2 density | 1/BUAPC, Urban population, Population density |

| PC1 prosperity<br>PC2 prosperity<br>PC3 prosperity<br>PC4 prosperity | Life expectancy, Infant mortality, GDP, HDI, Child mortality,<br>Alcohol consumption, Foreign-born population |
|---|---|
| PC1 disease<br>PC2 disease<br>PC3 disease<br>PC4 disease | Obesity (% age 20 and older),<br>Age 65+ CVD deaths,<br>Adults with hypertension (%),<br>Population with high cholesterol (%),<br>Population smoking (%),<br>Population with cardiovascular disease (%),<br>Population with diabetes%,<br>Population with cancer (%),<br>Population chronic kidney disease (%),<br>Population chronic obstructive pulmonary disease (%),<br>Population multiple chronic conditions (%),<br>Population physical inactivity (%), |
| PC1 ins.<br>PC2 ins.<br>PC3 ins. | No health insurance (% of_children_under_18),<br>No health insurance (% of_adults_ages_18–64),<br>No health insurance total population (%),<br>No health insurance black (%),<br>No health insurance native (%),<br>No health insurance Asian (%),<br>No health insurance Latino (%),<br>No health insurance white (%), |

**Table 2.** Grouping of variables and relation to PC.

Since different variables are expressed in different units and correspond to diverse scales, each variable in the dataset was standardized (the mean subtracted and divided by the standard deviation) before PCA. For each dataset, we retained as many PCs (starting from the most dominant one) as needed to (cumulatively) explain >85% of the data variance. It was inspected that PCs reasonably follow a normal distribution (as expected, based on the transformation of the original variables). Note that some of the initial variables did not satisfy our grouping criteria and thus do not appear in Table 2. They either have a distinct meaning from other variables (e.g., racial prevalence) or have a similar meaning, but do not exhibit a high correlation with the related variables (e.g., relative humidity RH2M, which does not correlate well with the other two humidity measures, QV2M and T2MDEW). These variables enter further analysis independently, i.e., together with PCs obtained after PCA on grouped variables.

### 2.7   LASSO regression

To complement the PCA feature selection, additional L1 regularization was done with Lasso (Hastie et al., 2009; Tibshirani, 1996). All input variables were standardized. Hyperparameter $\lambda$ (which controls the model complexity) was optimized through grid search on an exponential scale from numerical zero (OLS regression) to the value yielding the intercept-only model. Mean Squared Error (MSE) on the cross-validation testing set (200 repeats, 80-20 split) was taken as the loss function, and we chose the $\lambda_{1SE}$ as the simplest model still comparable to the optimal one (Krstajic et al., 2014). The final model was comprised of all the non-zero coefficients.

### 2.8   Elastic net regression

Elastic Net expands the Lasso regression with an L2 regularization and introduces a second hyperparameter $\alpha$ (Friedman et al., 2010; Hastie et al., 2009; Zou and Hastie, 2005). The same preprocessing was done for the input variables, after which the 2-dimensional grid-search with the same $\lambda$-scale as in Lasso, and the $\alpha$ linearly equidistant on the interval from 0 (Ridge regression) to 1 (Lasso regression) inclusive. Cross-validation was performed in the same way as for the Lasso regression, but each fold gave a distinct $(\alpha, \lambda)$ pair of hyperparameters. The final chosen value was the $(\alpha, \lambda)$ pair closest to the centroid of all the folds, and these hyperparameters were used to retrain the model on the whole dataset. Again, the final model was comprised of all the non-zero coefficients.

### 2.9 Random Forest and Gradient Boost

To avoid overfitting, the variables were preselected to exhibit significant correlations with $R_0$ (with a liberal threshold of P<0.1) by either Pearson, Kendall, or Spearman correlations. Cross-validation and hyperparameter selection for Gradient Boost (GBoost) and Random Forest (Breiman, 2001, 1996; Freund and Schapire, 1997; Friedman, 2001; Hastie et al., 2009) was done equivalently as for Lasso and Elastic net. For Gradient Boost, maximal number of splits, minimal leaf size, and learning rate were chosen through grid search, with the respective values: {1, 2, 3, 4, 5, 8, 16}; {1, 2, 3, 4, 5, 8, 16, 18} ; { 0.1, 0.25, 0.5, 0.75, 1}. For Random Forest, the grid values for the maximal number of splits and minimal leaf size were, respectively: {6, 12, 18, 22, 24, 26 30, 35}, {1, 2,…, 7}. In the ensemble, the number of trained decision trees was chosen to minimize Mean Square Error (MSE) on the testing set, for both methods. The obtained hyperparameters were used to retrain the models on the whole dataset, and predictor importance was estimated for both methods.

### 2.10 Model metrics

MSE for the testing data, averaged over all cross-validations, was used as a metric to compare the performance of different models. For easier interpretability, MSE values were scaled by those corresponding to the constant model (so that MSE of 1 corresponds to the constant model). To assess statistical significance with respect to the constant model, a t-test was applied to MSE values obtained through cross-validation.

## 3 Results

### 3.1 Extraction of $R_0$ and feature engineering

The $\log\big(D(\mathrm{t})\big)$ in the exponential growth regime for a subset of selected USA states is shown in Fig. 1. The linear dependence confirms that the progression of the epidemic in the early infection stage is almost perfectly exponential and is robustly observed for a wide range of USA states, while the same initial exponential growth was previously observed for a wide range of world countries (Notari and Torrieri, 2021; Salom et al., 2021). We exploited this exponential regime to infer $R_0$ as described in Methods, which we further use as our independent (response) variable.

Next, we transformed the variables so that their distribution became as close as possible to normal, and removed the outliers, followed by a grouping of the variables into subsets and performing PCA on these subsets, as detailed in Methods. The results of PCA are shown in Table 2, where each group of variables is related to their corresponding PCs in that table. For each variable group, we retained as many PCs as needed to explain more than 85% of the variability in the subset (standard threshold). To each of the PCs listed in Table 2, we assigned an intuitive name (e.g., PC1 prosperity, PC1 age) according to the set of variables from which they are formed.

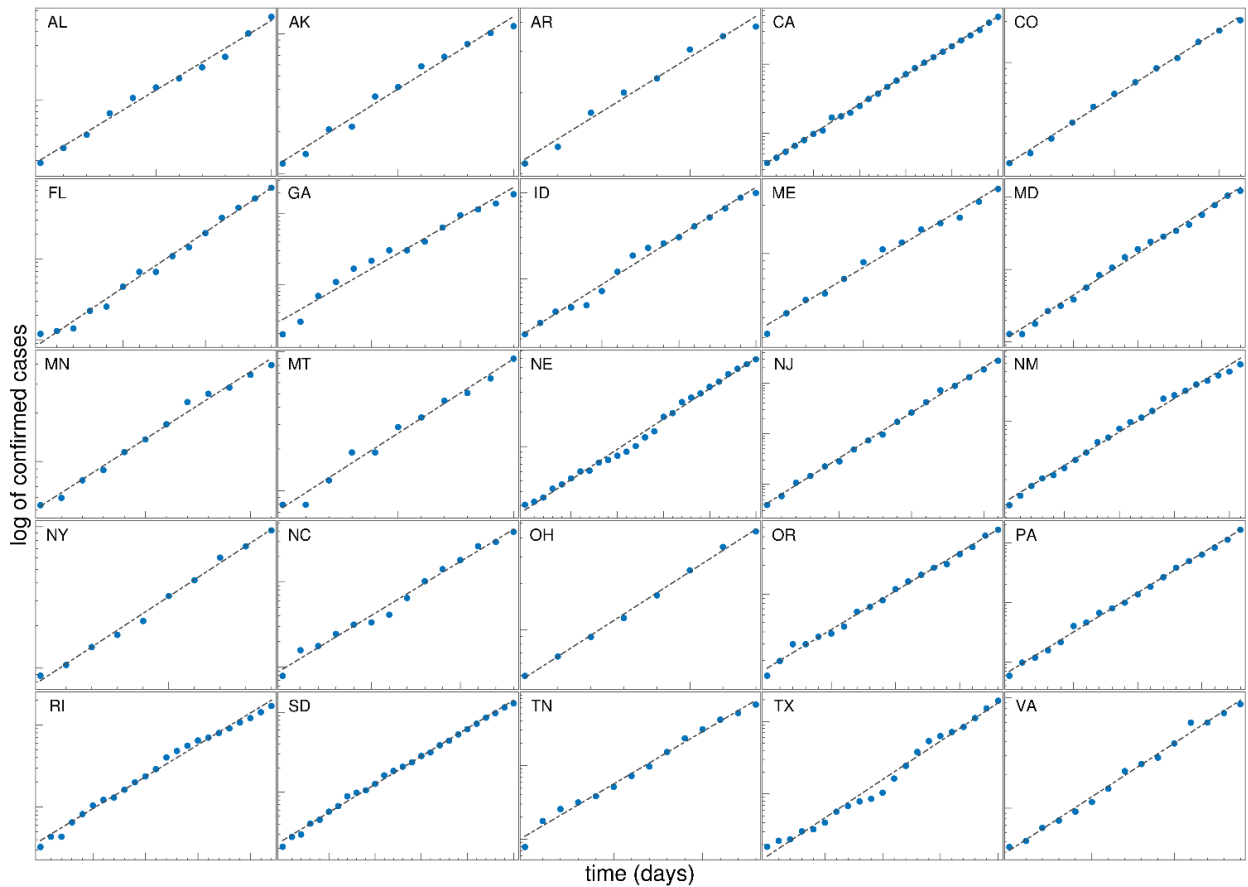

**Figure 1**. The time dependence of the detected cases for the different US states during the initial period of the epidemic is shown on a log-linear scale. The linear fit of log(D) shows that the spread of COVID-19 is well approximated by exponential growth in this phase. Values on axes are chosen differently for each state to emphasize the exponential growth phase. For each state, the start and end dates, the extracted slope $\lambda_+$, of the exponential regime, are given in Supplementary Table S1. Al – Alabama; AK – Arkansas; AR – Arizona; CA – California; CO – Colorado; FL – Florida; GA – Georgia; ID – Idaho; ME – Maine; MD – Maryland; MN – Montana; NE – Nebraska; NJ – New Jersey; NM – New Mexico; NY – New York; NC – North Carolina; OH – Ohio; OR – Oregon; PA – Pennsylvania; RI – Rhode Island; SD – South Dakota; TN – Tennessee; TX – Texas; VA – Virginia.

### 3.2    Feature extraction

We started from the basic assessment of the variable importance in explaining $R_0$, which are pairwise correlations. Note that these do not control for the presence of other potentially important variables but are a straightforward initial assessment of the relation with $R_0$. In Figure 2A, we show the Pearson correlation constant of the variables with $R_0$, where predictors with statistically significant correlations ($P<0.05$) are shown together with their correlation constants (represented by bars' heights) and statistical significance levels (indicated by stars). Somewhat surprisingly, we found that the highest correlation was with $PM_{2.5}$, with R~0.6 and P~$10^{-4}$. A large positive correlation between $R_0$ and $PM_{2.5}$ levels can also be observed from the scatter plot in Figure 2B. Additionally, several other variables exhibit statistically significant correlations with $R_0$, as indicated in Figure 2A. Note, however, that some of these variables are also significantly correlated with $PM_{2.5}$. Moreover, their correlation with $R_0$ and $PM_{2.5}$ is in the same direction (Figure 2C). Consequently, their significant correlation with $R_0$ may be, at least in part, due to their correlation with $PM_{2.5}$.

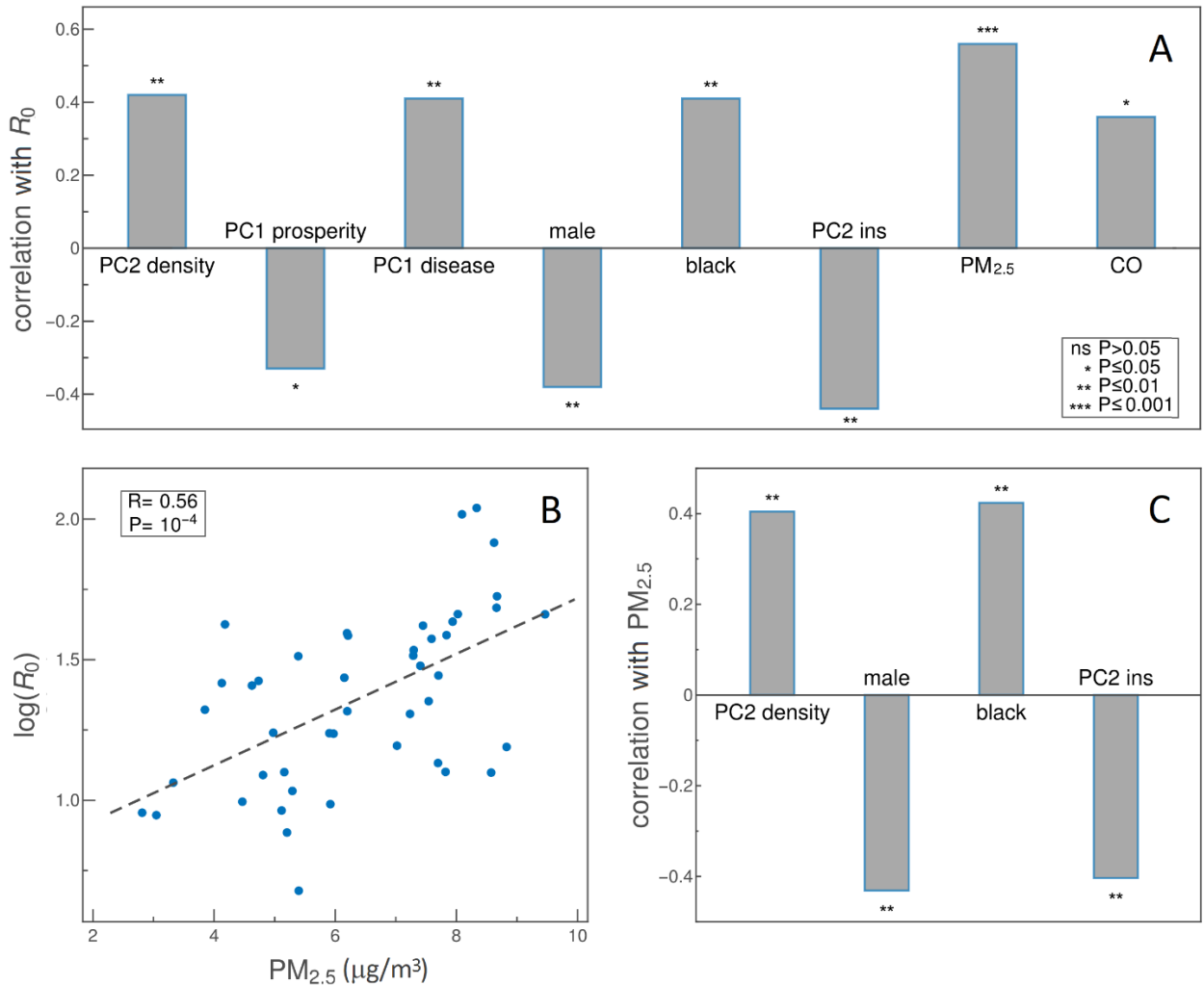

**Figure 2**. Pearson's correlations for relevant variables. A) Variables significantly correlated (P<0.05) with the basic reproduction number $R_0$ are shown. The bars' height indicates the value of Pearson's correlation coefficient (R on the y axis). B) Scatter plot of $R_0$ vs. $PM_{2.5}$. The dashed line shows linear fit. C) Person's correlations of variables in A) with $PM_{2.5}$. Variable names are indicated on the horizontal axis. Stars in bar plots represent the level of statistical significance, as indicated in the figure legend.

To partially address this, we performed an analysis that allows us to select the most important predictors from the set of correlated variables. Specifically, results of Lasso and Elastic net regressions are shown in Figures 3A and 3B. Both of these methods provide both regularization and the ability to select significant predictors through shrinking other coefficients to zero. Moreover, we standardized all the variables before using them in regressions, so that the absolute values of the regression coefficients provide estimates of relative importance of the selected variables. For each of the two methods, we performed repeated cross-validations, together with optimizations of hyperparameters, so that methods have maximal predictive power (minimal MSE) on the training set (see Methods for details). We obtained that the two methods are statistically highly significant compared to the constant model (P~$10^{-19}$ and $10^{-23}$, for Lasso and Elastic net, respectively). The predictive power of these methods is, however, only moderate, as can be seen for the obtained MSE values (MSEs are scaled, so that MSE of 1 corresponds to the constant model, which is not a large difference from 0.79 and 0.76, obtained by Lasso and Elastic net, respectively). Note, however, that the main purpose of these models was in feature selection, while predictability was improved through models employed in the next subsection.

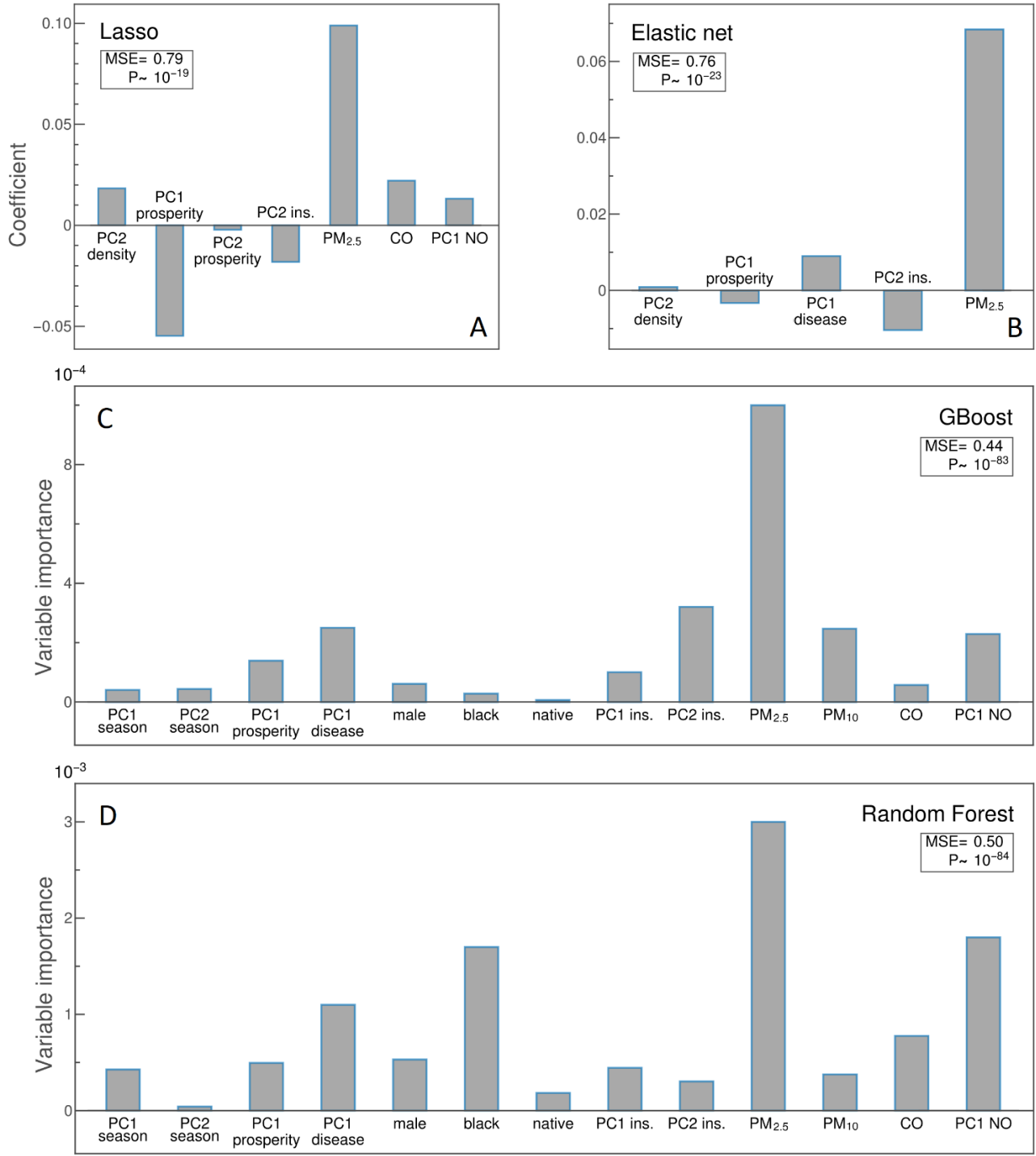

**Figure 3:** Values of regression coefficients in A) Lasso and B) Elastic Net regressions, respectively, where the bars' height corresponds to the coefficients' values for selected variables. Coefficients of all other variables are shrunk to zero (not shown) by the regressions. Variable importance in C) Gradient Boosting (GBoost) and D) Random Forest (RF) regressions, with the bars' height corresponding to estimated importance. Only variables with $P<0.1$ (according to either Pearson, Kendall, or Spearman correlations with $R_0$) are included in GBoost and RF regressions. MSE values are scaled to the constant value model and averaged over 200 cross-validations. $P$-values correspond to the statistical significance of obtained MSE's compared to the baseline model. Variable names are indicated on the horizontal axis.

From both Lasso and Elastic net, we again obtained that $PM_{2.5}$ was the most important predictor, positively affecting COVID-19 transmissibility (so that higher $PM_{2.5}$ leads to higher transmissibility). A similar trend was obtained for CO and PC1 NO (formed from NO2 and Nitrogen-oxides concentrations) – CO was also found to be significantly related with $R_0$ through pairwise correlations. Additionally, the population density (PC2 density) appears as an important predictor through both Lasso and Elastic net, though with smaller importance (regression coefficient), but consistently with pairwise correlations and with a tendency to increase transmissibility. Also, through all three approaches employed so far (pairwise correlations, Lasso, and Elastic net), we obtained that the higher state prosperity (PC1 prosperity) negatively influences $R_0$. Also, chronic diseases significantly influence (increase) $R_0$ as obtained by both pairwise correlations and Elastic net. Finally, PC2 ins., which is related to the fraction of the population (in particular Latinos) with medical insurance, also negatively correlates with $R_0$ (through all three methods). Interpretation of these dependencies is further addressed in the Discussion section.

### 3.3    Variable importance estimates

Our next goal was to assess variable importance and achieve better model predictability through methods that are considered state-of-the-art in machine learning for these types of problems. We employed two methods based on ensembles of weak learners (decision trees), in particular Gradient Boost and Random Forest. They are substantially different from Lasso and Elastic net employed in the previous subsection, as they do not assume linear dependence of the response from input variables (so-called non-parametric models). Consequently, their employment provided an independent check for the importance of PM$_{2.5}$ in explaining $R_0$. Our motivation was also to obtain better predictability of these models so that we can generate a quantitative estimate of pollution variation effects on $R_0$.

Two methods were implemented similarly to Lasso and Elastic net, i.e., model hyperparameters are optimized to achieve maximal predictability through repeated cross-validations (see Method for details). As these models (i.e., decision trees in general) are prone to overfitting, we performed a simple variable selection. That is, only variables with P<0.1 (according to either Pearson, Kendell, or Spearman correlations) were selected, resulting in 13 variables shown on the horizontal axes of Figures 3C and 3D, which were then used in further analysis. We obtained a much better predictive power for both Gradient Boost and Random Forest models (compared to regressions in the previous subsection) with MSE of 0.44 and 0.5, respectively, where these differences compared to the constant model (MSE=1) are statistically highly significant (P~$10^{-83}$ and $10^{-84}$, respectively).

Estimates of variable importance for both of these models are shown in Figures 3C and 3D. In both figures, the most prominent feature is PM$_{2.5}$, consistently with all other results obtained so far. Furthermore, PC1 disease and PC1 NO appear with moderate importance in both methods, where GBoost also emphasizes the importance of PC2 ins., which is all generally consistent with the analysis presented in the previous subsection. With respect to the pollution, the only difference is that PM$_{10}$ appears as moderately important in GBoost, while not selected by other models. Also, CO was selected by Random Forest as moderately important (consistent with the previous analysis) but does not appear as such in GBoost. Finally, the racial factor (in particular, fraction of black population) was selected as important by Random Forest (and also appeared as significant through pairwise correlations) but does not appear as important in GBoost. A possible interpretation of these findings is addressed in the Discussion section.

### 3.4    Quantitative estimate of pollution influence on $R_0$

As we obtained a reasonable model accuracy through both GBoost and Random Forrest, we were able to estimate how pollution variations (observed through different USA states) affect $R_0$. While we included a substantial number of variables (all that we managed to systematically assemble) in our analysis, these are of course not all the variables that can affect $R_0$, so we only aimed to provide rough estimates. Still, such an estimate is useful, as it provides the magnitude by which reasonably realistic changes in the pollution levels can affect $R_0$. For example, the new SARS-CoV-2 strain that was first detected in Great Britain (known as B.1.1.7, or more recently Alpha(Callaway, 2021)), which has, at the time of writing, become dominant in many other parts of the world, is estimated to lead to up to 1.9 increase in $R_0$ – this value can e.g. be compared with our estimated change due to pollution variations. To generate predictions for each of the analyzed states, we kept all other parameters fixed while changing the pollution values so that the changes corresponded to the actual values observed in all 46 states. In this way, the relative change in $R_0$, due to observed variations in pollution ($\Delta R_0/R_0$), was estimated, where $\Delta R_0$ corresponded to the difference between maximal and minimal estimated $R_0$ values.

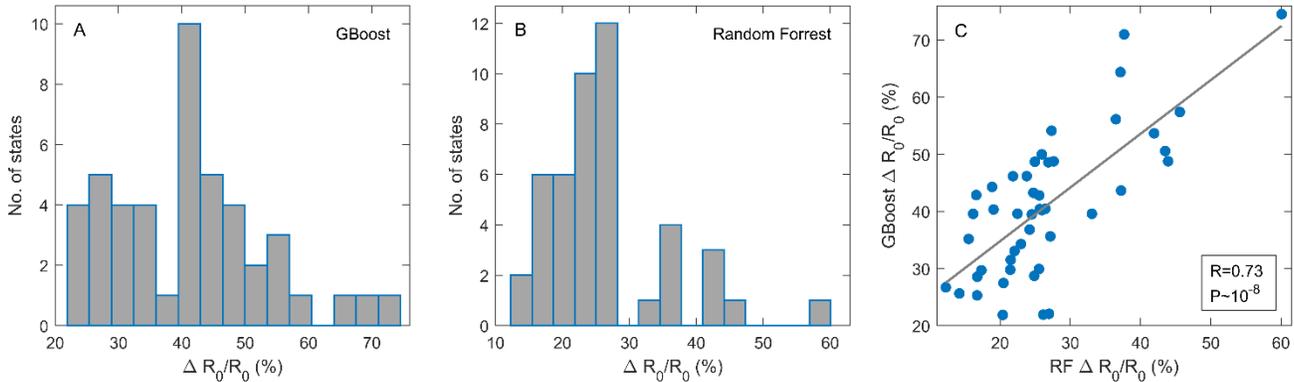

**Figure 4:** Relative change in $R_0$ due to pollution variations observed in USA states. For each state included in the analysis, $R_0$ was predicted for the range of pollution values observed throughout all other states. Relative variation in $R_0$ was estimated through both A) Gradient Boost (GBoost) and B) Random Forest (RF) regressions, with the models trained as in Fig. 3C) Scatter plot of $\Delta R_0/R_0$ predictions for GBoost and RF, with indicated Pearson's correlation coefficient and $P$-value.

The obtained results for $\Delta R_0/R_0$ for all analyzed states are shown as histograms in Figure 4A (GBoost) and 4B (Random Forest). For GBoost, a somewhat larger $\Delta R_0/R_0$, corresponding to the median of ~40% (and going up to ~70%), was obtained, while for Random Forrest, smaller values with a median of ~25% were estimated. This can e.g. be compared with $\Delta R_0/R_0$ of up to 90% for the Alpha strain (Davies et al., 2021) so that estimated changes due to pollution variation are smaller but still substantial. Finally, as the two histograms are somewhat different, in Figure 4C we directly test the consistency of their $\Delta R_0/R_0$ predictions. It can be seen that they are well consistent, with reasonably high correlation (R=0.73 and P~10⁻⁸). Note that these two methods are independent and substantially different (though both based on ensembles of decision trees), so differences in their predictions are expected.

## 4    Discussion

Figures 2 and 3 reveal the main result of the paper: $PM_{2.5}$ pollution is, throughout our analysis, consistently singled out as the main driver behind SARS-CoV-2 transmissibility in the US. This result was obtained through both pairwise correlations of variables with $R_0$, and by the applied machine learning approaches.

The association of the $PM_{2.5}$ pollution with the rate of COVID-19 spread *per se* is not a novel result (Gujral and Sinha, 2021; Gupta and Gharehgozli, 2020; Kolluru et al., 2021; Lorenzo et al., 2021; Maleki et al., 2021; Stieb et al., 2020). However, the existing studies had several methodological limitations (Anand et al., 2021; Bontempi, 2021; Bontempi et al., 2020; Villeneuve Paul J. and Goldberg Mark S., 2020), outlined in the Introduction, that we here tried to address. Moreover, previous studies in the USA obtained non-consistent reports on pollution relevance, underlying the importance of more extensive modeling and statistical learning approaches that we employed here (Allen et al., 2021; Gupta and Gharehgozli, 2020; Luo et al., 2021).

First of all, by explicitly taking into account the infection dynamics, i.e., the model-based estimate of $R_0$ as SARS-CoV-2 transmissibility measure (instead of, for example, considering case counts) we addressed a number of common shortcomings of studies with a similar goal: $R_0$ obtained in this way (as it depends only on the curve exponent and is thus scaling invariant) is prone neither to underreporting bias nor to errors due to differences in testing policies (Villeneuve Paul J. and Goldberg Mark S., 2020); since we concentrate only on the initial period of the local epidemic, our results do not suffer from the problem of comparing different stages on the epidemic curves, are not influenced by the existence of multiple epidemic peaks nor by the later appearance of multiple virus strains, and are unaffected by social measures which alter dynamic only later (Bontempi, 2021; Villeneuve Paul J. and Goldberg Mark S., 2020); our approach does not rely on time series and thus avoids the related methodological difficulties (Villeneuve Paul J. and Goldberg Mark S., 2020). Next, as our inferences are not based simply on mutual correlations of variables alone, but we also robustly obtain the same main conclusion by employing four different machine learning techniques, including those that can account for potentially highly non-linear dependences of $R_0$ on predictors. Consequently, common objections to statistical analysis (Bontempi et al., 2020; Villeneuve Paul J. and Goldberg Mark S., 2020) do not apply here. Furthermore, by taking into account 74 diverse predictors covering a broad scope of potentially relevant factors, we avoid the lack of multidimensionality and a bias that may result from considering only a narrow class of variables – problems otherwise observed in many similar studies (Bontempi, 2020b; Bontempi et al., 2020). With regards to that, we note that our study was initially conceptualized to explore which parameters, from a large collected set, had the most influence on the spread of the SARS-CoV-2 virus in the USA (without initial bias towards pollution). As our preliminary results singled out air pollution as the major predictor of COVID-19 transmission speed, this motivated us to put the pollution variables in the spotlight of this research, trying also to differentiate which types of pollution mostly contribute to the transmission of COVID-19.

As several limitations still remain in our study, the observed association between $PM_{2.5}$ pollution and COVID-19 cannot be yet taken to guarantee the existence of a causal relation. Even with the use of advanced statistical learning methods, it is difficult and not always possible to disentangle the effects of strongly correlated variables. As we will further discuss below, it is particularly problematic to differentiate between the independent effects of pollution and the indirect effects of factors related to economic and racial disparities, which often go hand in hand in the USA (Chakraborty, 2021). Another problem is to select a proper proxy (or proxies) for the frequency of human interactions in a given society, as there is little doubt that the human-to-human mode of transmission is most dominant in COVID-19. In this context, some authors (Bontempi, 2020b; Bontempi et al., 2020; Cartenì et al., 2020; Guo et al., 2021) rightfully emphasize the importance of properly assessing the mobility of the considered population, and suggest possible proxies: from specific measures of economic relations and commercial exchanges to taking into account the number of job seekers/investors and analysis of public transportation statistics. Presently, we have taken into account only basic measures of economic prosperity that are expected to indirectly but highly correlate with mobility and frequency of human to human interactions: human development index, gross domestic product per capita, life expectancy, infant/child mortality, and foreign-born population. While it is not easy to identify and find further variables that could properly reflect these factors and yet be available, in a systematic and unified way, across all studied regions, there is certainly room for methodological improvement in this respect.

Another methodological limitation that cannot be easily overcome is the potential difference between indoor and outdoor air pollution. This is of obvious relevance since it is estimated that people, on average, spend 80-90 percent of their time indoors (Noorimotlagh et al., 2021b). In the absence of systematic data sources on indoor pollution, our conclusion must rely on a reasonable assumption that indoor and outdoor pollutions are, in general, highly correlated, as is illustrated in (Harbizadeh et al., 2019). The unavoidable trade-off between choosing a scope of analysis that exhibits extreme levels (and variations) of air pollution on one side, and the need for uniformity of other parameters on the other side – that we settled by choosing the USA dataset – presents an additional limitation, considering that pollution values in the USA are generally not high, and certainly below serious health-hazard levels. (The values are far below the levels investigated in the COVID-19 context in some other locations: for example, in a study done in Bangkok (Sangkham et al., 2021) the authors reported much higher $PM_{2.5}$ values but had to face severe methodological limitations of the sorts discussed above.) Despite these remaining limitations of our research, we believe that this work presents substantial progress in terms of methodology and reliability of the obtained results. It thus establishes the link of $PM_{2.5}$ pollution with COVID-19 transmissibility much more firmly than the previous studies and provides further motivation for research in this direction.

Since this study suggested a direct relation between pollution and COVID-19 transmissibility, we finally provided a quantitative estimate of the established connection in Figure 4. We estimated that varying the pollutant levels (specifically, levels of $PM_{2.5}$, $PM_{10}$, CO, and $NO_2$, which enter Random Forest and Gradient Boost methods), where changes in $PM_{2.5}$ levels are by far the most important, makes a difference of ~30% in terms of the $R_0$ values. While this is smaller compared to reproduction number changes due to the appearance of new highly infective strains (estimated to increase $R_0$ for up to ~90% higher) (Davies et al., 2021), it is still sizable, and clearly illustrates the potential importance of $PM_{2.5}$ in modulating the virus transmissibility. For example, in an exponential regime of infection progression (c.f. Eq. (1.7) in Methods) lasting for ~10 days (a typical period in which exponential growth is observed for the USA states), and with typical parameter values, such difference would lead to two times larger number of infected, and (at least) equal proportion of lost human lives. Aside from increasing transmissibility, an additional (and largely independent) effect of larger pollutant levels is the potentially increased COVID-19 mortality (due to health hazards of pollution), as suggested by several studies (Luo et al., 2021; Pozzer et al., 2020; Wu et al., 2020). Overall, this underscores the importance of reducing pollutant levels in the epidemiological context, along with other established non-pharmaceutical measures (Abboah-Offei et al., 2021; Anand et al., 2021; Bontempi, 2021).

While we obtain that $PM_{2.5}$ pollution is the dominant predictor of virus transmissibility, our results also identify the relevance of other factors. First, a few other pollutants are also selected through our analysis, most notably $NO_2$ and its related nitrogen oxide derivatives (where its particularly high importance was assigned by the Random Forest method, see Fig. 3D), and to some extent CO and $PM_{10}$. These results are partially in line with findings that several pollutants, more precisely particulate matter (Comunian et al., 2020; Sagawa et al., 2021), but also $NO_2$ (Paital and Agrawal, 2020), cause overexpression of ACE-2 in respiratory cells, thus increasing the likelihood of infection. This is not the only potentially relevant mechanism, as some studies point to the prolonged exposure to pollutants as a cause of a general weakening of the immune system (Glencross et al., 2020; Qu et al., 2020). However, the relatively low importance of NO and CO pollutants that we obtained speaks in more in favor of the hypothesis that PM pollution, by binding to virus droplets, mechanically facilitates SARS-CoV-2 spread through the air - both extending the range of virus diffusion and allowing its direct transport into deeper pulmonary regions (Qu et al., 2020). This suggested mechanism of the pollution-to-human mode of transmission should be seen in the light of substantial evidence for COVID-19 airborne transmission via aerosols (Anand et al., 2021; Kenarkoohi et al., 2020; Noorimotlagh et al., 2021b, 2021a) and of established positive correlation between the concentration of certain pathogens in air and PM pollution (Chen et al., 2010; Harbizadeh et al., 2019). The fact that we were here considering short-term (acute) pollution values precisely in the initial days of the outbreak, and the fact that pollution levels in the US are well below serious health hazards, are also in favor of this mechanistic interpretation of the pollution-COVID-19 link, rather than of the explanation via general adverse effects of pollution on the immune system. On the other hand, the inferred large difference in the influence of $PM_{2.5}$ and $PM_{10}$ particles may be understood through the difficulty of particulate matter larger than 5 µm to reach ACE2 receptors located in type II alveolar cells (Copat et al., 2020; Zhu et al., 2020). It should be noted that our study is not the only one suggesting a substantial difference between the effect of $PM_{2.5}$ and $PM_{10}$ particles on the spread of COVID-19 (Copat et al., 2020; Lorenzo et al., 2021; Zhu et al., 2020).

Another factor (unsurprisingly) related to the susceptibility of an organism to infections, is the presence of different comorbidities and, in general, any diseases that could potentially compromise the immune system (Allel et al., 2020; Coccia, 2020; Liu et al., 2020). Indeed, all applied analysis methods except for Lasso find the prevalence of chronic diseases in the population (i.e., its dominant principal component PC1-disease) to be an important $R_0$ predictor.

Additionally, our applied methods also identify a group of three mutually interrelated factors: the dominant PC reflecting the overall prosperity of the state (PC1 prosperity), the percentage of the black population, and the PC2 insurance component (this component effectively reflects the insurance coverage among the Hispanic population). Our recent study of the effects of various demographic and weather parameters on the spread of COVID-19 based on the data from 118 world countries (Marko Djordjevic et al., 2021) also pointed to the essential role of the country's prosperity, but we note a disagreement in the sign of the correlation: whereas, worldwide, the more developed countries suffered from higher COVID-19 expansion rates, data on US states show an opposite trend - wealthier and more developed areas of US on average seem to exhibit lower $R_0$ values (Gupta and Gharehgozli, 2020). However, this difference may be expected: on the global level, there are substantial variations in the development level between countries, and this level effectively becomes a proxy for the frequency of social contacts (reflecting business and cultural activity, population mixing due to work/education, international travel, etc.) (Bontempi, 2020b; Bontempi et al., 2020; Gangemi et al., 2020). On the other hand, US states have highly developed societies and the dominant effect of these more subtle differences is likely different: within this prosperity range, the better off population has more means to prioritize and practice precautionary behavior (e.g., have professions that require less physical contacts, fewer comorbidities, healthier lifestyle, higher awareness of the infection risks, etc.). Furthermore, compared to the global analysis, we note that air pollution also played a role in that study, though a less prominent one, via a principal component that turned out to encapsulate also other measures of unhealthy living conditions and lifestyle. While the influence of PM2.5 on COVID19 transmission should, of course, exist everywhere and cannot be effect unique to the territory of the USA, we note that this influence is much more difficult to observe when considering more diverse areas/populations, as it might be overshadowed by more dominant factors.

The COVID-19 pandemic has also emphasized a specific racial aspect of healthcare disparities. The correlation between the percentage of the black population and $R_0$ observed in our data (Figure 2A), as well as the results of the Random Forrest regression method (Figure 3D), agree with the already established conclusion that the black minority is by far overrepresented not only among COVID-19 fatalities (Luo et al., 2021; Wu et al., 2020) but also among the total infected population (Chakraborty, 2021). Another relevant factor is the health insurance coverage (PC2 insurance), which consistently through our analysis shows that COVID-19 infection is spreading faster among people without medical insurance (Figure 2 and Figure 3). Both the percentage of the black population and the prevalence of insurance coverage are significantly correlated with pollution, in particular with $PM_{2.5}$, as can be seen in Figure 2C (curiously, our data do not show such correlation with the PC1 prosperity component). Further complicating this relation of poverty, pollution and COVID-19, are the findings that indicate the importance of high quality and well maintained artificial ventilation (which is not equally affordable to everyone) in reducing indoor pollution with possible consequent effects on COVID-19 transmission (Harbizadeh et al., 2019; Noorimotlagh et al., 2021b). It has been already argued that the influence of factors related to a more economically disadvantaged population (overrepresentation of minorities, absence of medical insurance,…) is inherently hard to disentangle from the effects of pollution (Chakraborty, 2021). While this standpoint is also in part supported by our analysis, we also note that $PM_{2.5}$ consistently appeared with much larger importance through all analyses compared to these economically disadvantaged factors (Chakraborty, 2021; Stieb et al., 2020). In this sense, based on our results, it seems more plausible to associate $PM_{2.5}$ (rather than these other factors) with $R_0$ changes.

It is also interesting to consider which parameters did not show up as important in our results. The absence of seasonal principal components from the final sets of significant predictors may indicate that the importance of the weather parameters such as temperature, UV radiation, and humidity on the SarS-CoV2 transmission is lesser than commonly assumed. While there are substantial arguments that high temperatures and humidity levels should suppress virus transmission (Byun et al., 2021; Fu et al., 2021; Noorimotlagh et al., 2021a; Notari, 2021; Sarkodie and Owusu, 2020), the literature is not fully unison on this conclusion, with some even reporting the opposite effects (Kolluru et al., 2021; Lorenzo et al., 2021; Sangkham et al., 2021). The results presented here seem to side with some authors who disagree that weather factors bear a significant influence on the course of the COVID-19 epidemic (Wang et al., 2021). One should however note that variations of meteorological factors are much larger on a global scale, where indeed we find out a larger significance of these factors (Salom et al., 2021). Another somewhat surprising conclusion is the moderate significance of the population density that we obtain. While there is a significant correlation of PC2 density component with $R_0$, it further appeared significant only in Lasso regression, and even there with not a quite high coefficient. While in disagreement with common expectation and some studies (Chakraborty, 2021), this is however in line with several other studies, that also didn't assign a high significance to population density (Carozzi et al., 2020; Hamidi et al., 2020; Pourghasemi et al., 2020; Rashed et al., 2020). Rather than interpreting such an outcome simply as the irrelevance of population density, we, as already argued in (Salom et al., 2021), see it as an indication that more subtle measures of density (that would more accurately reflect effective proximity of individuals in everyday scenarios) are needed.

## 5    Conclusion and outlook

Starting from 74 initial parameters and by using five different analysis approaches, we obtained the results that robustly select $PM_{2.5}$ pollution as a major predictor of SARS-CoV-2 transmissibility in the USA. Using $R_0$ as a transmissibility measure and non-linear dynamics to extract its values for different USA states, these results are largely insensitive to the differences in the state policies. The obtained large quantitative estimate of the magnitude of the $PM_{2.5}$ effect on virus transmissibility may be intuitively unexpected and is not that far from estimated differences in transmissibility caused by virus mutations.

The main issue to be addressed in future studies is that of causality, i.e., disentangling the effects of pollution from those of socio-demographic factors with which it is correlated. This clearly cannot be achieved through studies with low resolution, such as the one employed here, despite using sophisticated statistical (machine) learning methods and studiously taking into account the infection progression dynamics. Carefully crafted, and high-resolution, longitudinal epidemiological studies may be a way forward in this regard. The results obtained here, and by other similar studies, may provide a basis for these high-resolution studies, particularly in terms of factors that should be considered, their expected relative importance, and the magnitude of the effects that may be expected.

## 6    Conflict of Interest

The authors declare that the research was conducted in the absence of any commercial or financial relationships that could be construed as a potential conflict of interest.

## 7    Author Contributions

MarD, IS and MagD conceived the research. The work was supervised by MarD, IS and AR. Data acquisition and supplementary material by OM, MT and DZ. Code writing and data analysis by OM, MarD, DZ, SM. Figures and tables made by OM, DZ, MT and MagD. A literature search by AR and SM. Result interpretation by MarD, MagD, IS, AR and SM. Manuscript written by MarD, IS, AR, MagD, OM and MT.

## 8    Data Availability Statement

Data is provided in the Supplementary material.

## 9    Funding

This work was partially supported by the Ministry of Education, Science and Technological Development of the Republic of Serbia.